\documentclass[conference]{IEEEtran}
\IEEEoverridecommandlockouts

\usepackage{cite}
\usepackage{amsmath,amssymb,amsfonts}
\usepackage{algorithmic}
\usepackage{graphicx}
\usepackage{balance}
\usepackage{textcomp}
\usepackage{xcolor}
\usepackage{url}

\def\BibTeX{{\rm B\kern-.05em{\sc i\kern-.025em b}\kern-.08em
    T\kern-.1667em\lower.7ex\hbox{E}\kern-.125emX}}
\begin{document}

\title{Beyond Gaze Overlap: Analyzing Joint Visual Attention Dynamics Using Egocentric Data}

\author{
\IEEEauthorblockN{Kumushini Thennakoon, Yasasi Abeysinghe, Bhanuka Mahanama, Vikas Ashok, Sampath Jayarathna}
\IEEEauthorblockA{\textit{Department of Computer Science}, 
\textit{Old Dominion University}, Norfolk, VA, USA \\
kumushini@cs.odu.edu, yasasi@cs.odu.edu, bhanuka@cs.odu.edu, vganjigu@cs.odu.edu, sampath@cs.odu.edu}
}
 

\maketitle

\begin{abstract}
Joint visual attention (JVA) provides informative cues on human behavior during social interactions. The ubiquity of egocentric eye-trackers and large-scale datasets on everyday interactions offer research opportunities in identifying JVA in multi-user environments. We propose a novel approach utilizing spatiotemporal tubes centered on attention rendered by individual gaze and detect JVA using deep-learning-based feature mapping. Our results reveal object-focused collaborative tasks to yield higher JVA (44-46\%), whereas independent tasks yield lower (4-5\%) attention. 
Beyond JVA, we analyze attention characteristics using ambient-focal attention coefficient $\mathcal{K}$ to understand the qualitative aspects of shared attention. Our analysis reveals $\mathcal{K}$ to converge instances where participants interact with shared objects while diverging when independent. 
While our study presents seminal findings on joint attention with egocentric commodity eye trackers, it indicates the potential utility of our approach in psychology, human-computer interaction, and social robotics, particularly in understanding attention coordination mechanisms in ecologically valid contexts.

\end{abstract}


\section{Introduction}

The ability to understand where and how individuals direct their visual attention during social interactions provides details about the underlying cognitive processes, social cues, and coordination strategies that shape mutual understanding \cite{schneider-2021-shared-gaze-visualization}. Observing moments of Joint Visual Attention (JVA) during different social interactions between two or more individuals is a well-established method in developmental psychology \cite{baldwin2014, tomasello1986joint}. JVA is used in language acquisition research with children and in the early detection of autism \cite{charman2003joint}. Beyond psychology, JVA is widely used in various domains, including human-computer interaction \cite{mahanama2023disetrac,abeysinghe2024disetrac}, education, neuroscience \cite{duncan1998converging}, and social robotics \cite{scassellati1998imitation}. Traditional methods relied on qualitative approaches (e.g., gaze following \cite{cui2024gaze}, pointing) to analyze JVA. However, with advances in eye-tracking technology in conjunction with wearable eye-tracking, it has become easier for researchers to collect accurate gaze data, making the study of JVA more efficient and objective \cite{schneider2024, becker2021}.


Wearable eye-tracking technology has emerged as a powerful tool to capture egocentric visual behavior in real-world environments \cite{Macinnes299925, bock2024wear, wang2024egocentric, alletto2016exploring}. These devices provide a mobile and unobtrusive means of collecting multi-modal data from the wearer's point of view. Although much research on JVA has been conducted in controlled laboratory settings, only a few studies have focused on naturalistic, dynamic environments with more inherited constraints when studying the visual attention of multiple individuals. 

In this work, we present an analysis of JVA using egocentric video data and gaze data collected using wearable eye-tracking glasses. By leveraging the mobility and naturalistic recording capabilities of the devices, we aim to explore how JVA can be detected and interpreted in real-world social interactions. Our research contributes to ongoing efforts to bridge the gap between controlled experimental designs and the complexity of real-world behavior. We used the advance gaze metric ambient-focal attention coefficient $\mathcal{K}$ \cite{mahanama2022eye,krejtz2016discerning,jayawardena2025real} to observe the attention patterns of the participants. Data were obtained from the publicly available Aria everyday activities dataset \cite{lv2024aria} released by Meta as part of their larger vision for developing Augmented Reality (AR) and Artificial Intelligence (AI) technologies. This dataset was collected using Project Aria glasses \cite{engel2023project} during various daily activities such as cooking, watching television, making coffee, and engaging in conversations.

\section{Related work}


JVA has been studied across multiple disciplines with varying approaches and objectives. This section reviews relevant literature that informs our approach to analyzing joint visual attention in everyday activities using egocentric data.

Traditionally, JVA has been a central concept in developmental psychology, where it has been extensively used to understand typical development patterns and identify developmental conditions like autism spectrum disorder \cite{corkum2014development, baldwin2014, charman2003joint}. These studies relied mainly on qualitative observational methods \cite{yu2013joint, bradley2023qualitative} to assess when and how individuals coordinate their attention to the same object or event. While qualitative approaches provided valuable insights into the social aspects of attention, they lacked precision in measuring the exact properties of visual attention.

Advancements in eye-tracking technology have enabled researchers to measure JVA with higher precision. Schneider et al. \cite{schneider-2021-shared-gaze-visualization} proposed quantifying JVA in collaborative environments based on gaze overlapping. However, this simplified approach tends to reduce the complex nature of joint attention to a binary measurement (overlap or no overlap), overlooking the rich temporal and qualitative characteristics of shared attention. Our work aims to provide details beyond gaze overlap metrics to develop a more sophisticated approach to observing and understanding JVA.

Several studies have explored the identification of JVA using egocentric data from wearable cameras or eye trackers \cite{peters2021modeling, huang2020ego, kera2016discovering}. However, these studies have focused only on identifying instances of JVA but analyzing the attention behaviors of multiple participants simultaneously. Our research extends previous work by not only identifying moments of joint attention but also analyzing patterns and quantifying the percentage of JVA in everyday activities involving two participants.

Object detection has emerged as a promising approach for identifying JVA in egocentric settings \cite{peters2021modeling, tu2023joint}, particularly in controlled environments with limited sets of objects. However, this approach faces significant challenges when applied to everyday activities where participants interact with numerous diverse objects that may not be easily recognizable by standard object detection algorithms. Moreover, our primary goal is not to develop new methods for identifying JVA but rather to employ established methods of identification while focusing on deeper analysis of attention patterns.

Researchers have attempted to localize attention by analyzing the angle of direction from multiple video sources\cite{park20123d,jayawardena2021automated}. While this approach provides valuable spatial information, it often lacks the precision offered by eye-tracking data. Eye tracking provides more accurate information about the exact focus of attention \cite{mahanama2022eye}, which is particularly important when analyzing fine-grained attentional behaviors in everyday activities where the objects of interest may be in close proximity to one another.

Our method builds upon the idea of work by Kera et al. \cite{kera2016discovering} where they create a spatiotemporal tube by extracting regions of interest around the gaze position from egocentric video frames and calculating similarity between the defined tubes to identify instances of JVA. However, we extend this approach by incorporating advanced gaze measures to analyze the identified instances of JVA. This additional analytical step allows us to move beyond simple identification to understand the patterns of joint attention in everyday activities captured through egocentric recordings.
By addressing these research gaps, our work contributes to a more comprehensive understanding of JVA in naturalistic settings, with implications for both theoretical models of social attention and practical applications in fields such as education,

\section{Methodology}

\begin{figure*}
    \centering
    \includegraphics[width=\textwidth]{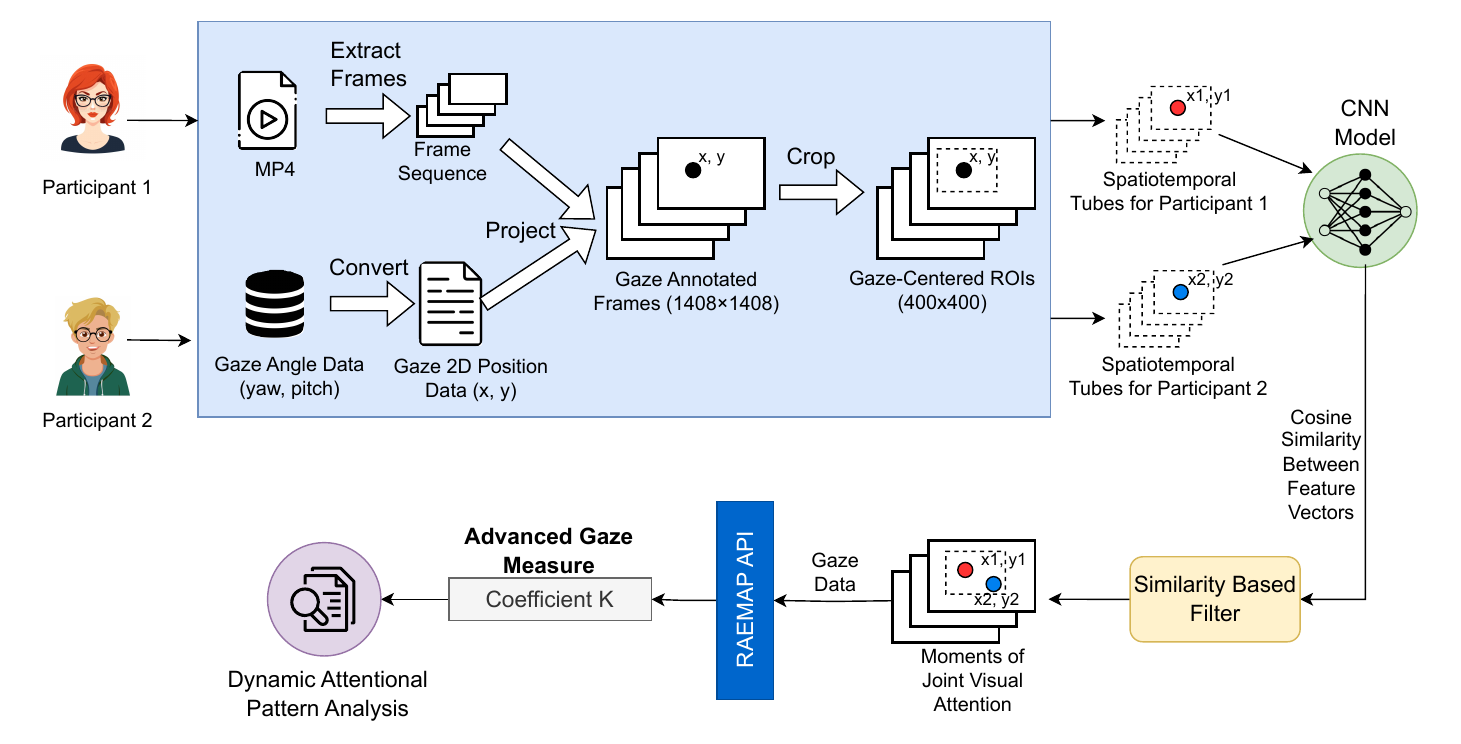} 
    \caption{\textbf{Processing pipeline for analyzing JVA between dyads in an egocentric setting}. We use egocentric video and gaze data from each participant in dual-participant activities in the Aria Everyday Activities Dataset \cite{lv2024aria}. Image frames are extracted from the video stream, and corresponding 2D gaze points are projected onto each frame. For each gaze-annotated frame, a 400×400 pixel region centered at the gaze point is cropped. Time-synchronized pairs of these gaze-centered regions (spatiotemporal tubes) are compared using a deep-learning model. Using a similarity threshold, we filtered the moments of joint visual attention. Gaze data from these moments are then processed through the REAMAP API \cite{jayawardena2019eye} to compute Coefficient $\mathcal{K}$, which is analyzed to assess dynamic attention characteristics.}
    \label{fig:diagram}
\end{figure*}

\subsection{Dataset}

For this study, we utilized the publicly available Aria Everyday Activities dataset \cite{lv2024aria}, which provides egocentric recordings of participants engaged in common daily activities. This dataset contains time-synchronized video data captured from a first-person perspective as individuals perform routine tasks such as cooking, making coffee, watching television, etc. In addition to video, the dataset provides multi-modal sensor data recorded using Project Aria glasses \cite{engel2023project}, including per-frame 3D eye gaze directional vectors. The dataset is particularly valuable for our research as it includes dual-participant recordings where two individuals interact in the same environment while both wearing Project Aria glasses. This synchronized dual-perspective video data along with gaze data allows us to examine JVA as it naturally occurs between pairs of participants during routine interactions. In this study, we used the dual-participant recordings from the dataset to analyze how individuals coordinated their visual attention across various everyday tasks, enabling the investigation of JVA patterns in real-world settings.

The Aria Everyday Activities dataset comprises 143 sequences of daily activities recorded by multiple participants across five geographically diverse indoor locations. However, most of these recordings involve only a single participant or missing recordings from one of the two participants. Since our focus is on dual-participant interactions, we excluded such data, resulting in a subset of 10 recordings. Among these, 5 recording sessions were further excluded due to highly dynamic motion, face-to-face conversations without shared object interactions, or a lack of common visual perspectives. After this filtering process, we retained 5 recordings for analysis. In the selected sessions, five participants, paired in varying combinations, took part in the five distinct activity sessions.

The data for each dual-participant task were processed through the pipeline shown in Figure \ref{fig:diagram} to facilitate the analysis of JVA.

\subsection{Extracting Regions of Interest (ROIs) around the Gaze Position}

\begin{figure*}
    \centering
    \includegraphics[width=\textwidth]{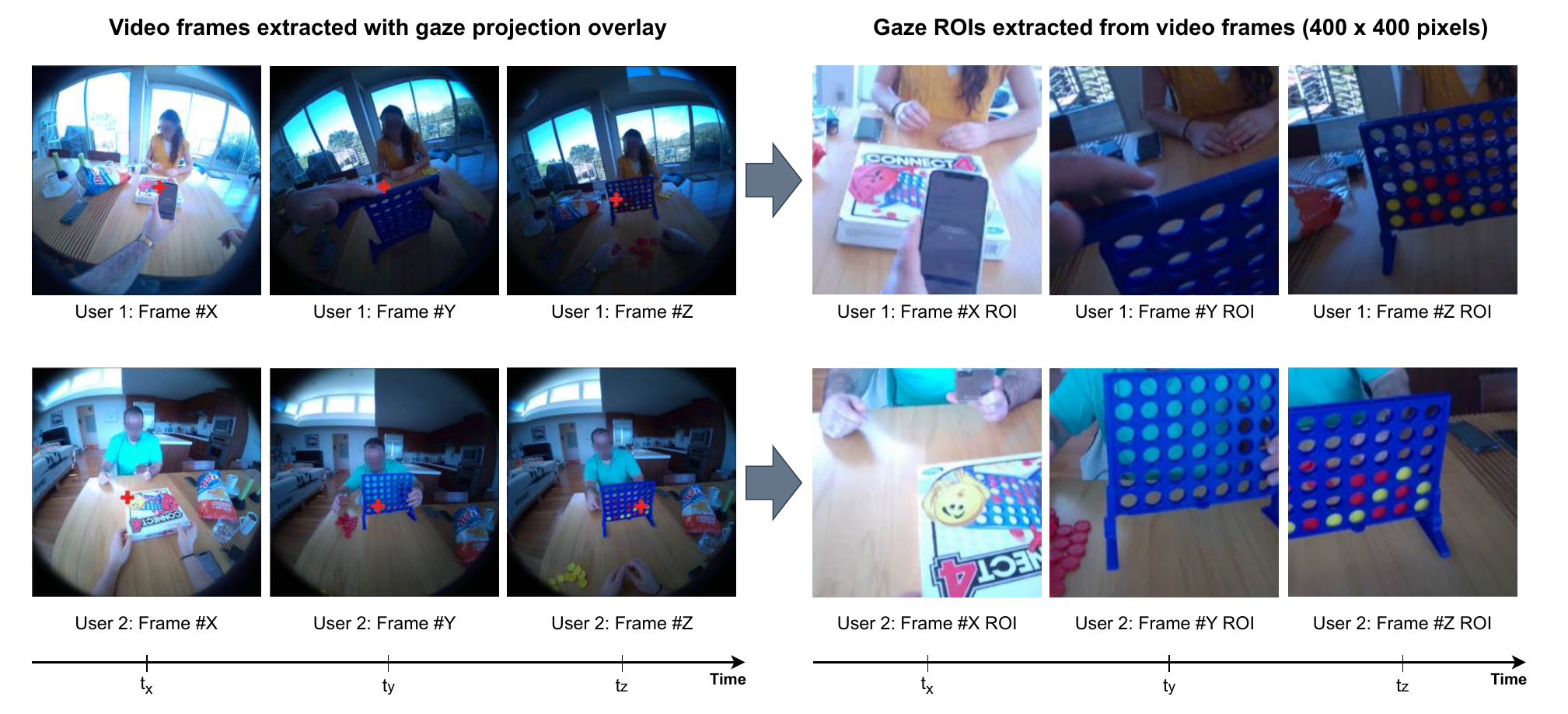} 
    \caption{\textbf{Extracting regions of interest (ROIs) around the gaze position from video frames.} The images on the left show the extracted frames from videos for both participants along the time. The images on the right show the 400$\times$400 area around the gaze position extracted from each video frame, making it a spatiotemporal tube \cite{kera2016discovering}.}
    \label{fig:gaze_rois}
\end{figure*}

To analyze JVA between dual participants, we needed to identify their shared visual perspective during the task. We began by decomposing the egocentric video streams into individual frames. Using Project Aria tools \cite{lv2024aria}, we converted the per-frame 3D eye gaze directional vectors into 2D gaze coordinates and projected these coordinates onto the corresponding video frames.

To detect moments of shared gaze, we initially compared the full image frames between participants. However, this approach yielded low similarity scores due to the presence of many unrelated objects in the scene. To better capture the visual content near the gaze point, we adopted a method based on spatiotemporal tubes around points of gaze \cite{kera2016discovering}, which allows for extracting and comparing the specific regions attended to by each participant.

From the gaze-annotated frames, we extracted a 400$\times$400 pixel region centered on each participant’s gaze point (see Figure \ref{fig:gaze_rois}). This window size of 400 pixels was empirically chosen to cover more than 25$\%$ of both the width and height of each frame (resolution: 1408$\times$1408), effectively capturing the focal area of visual attention while reducing peripheral visual noise. Assembled over time, these gaze-centered ROIs formed what we refer to as spatiotemporal tubes, enabling a more precise analysis of shared attention.



\subsection{Calculating Similarity Between Spatiotemporal Tubes}


In this study, we assess the visual similarity between two sets of image frames, referred to as spatiotemporal tubes using a deep learning-based feature extraction approach. Each frame is processed with a pre-trained ResNet-50 convolutional neural network (CNN) \cite{he2016deep}, a model widely adopted for image recognition tasks. Rather than using the network for classification, we remove its final classification layer and use the remaining layers to extract deep feature representations that capture high-level visual patterns from each image. 

These feature vectors are then compared using cosine similarity, a metric that quantifies the angular distance between two vectors in high-dimensional space. A higher cosine similarity score indicates greater visual similarity between the frames. By aligning frames based on their temporal positions, we compute similarity scores between image pairs across the two spatiotemporal tubes. This allows us to analyze how closely the visual content is aligned between two users or perspectives over time.

\subsection{Moments of Joint Visual Attention and Dynamics of Coefficient $\mathcal{K}$}

To identify moments of JVA between participants, we established a similarity threshold of 0.7 for comparing the spatiotemporal tubes. This threshold was empirically determined through careful manual inspection and comparison of various cases of joint attention and independent viewing. Different threshold values were evaluated, and the final value was selected based on their consistency in distinguishing similarities. When the similarity score between the tubes of both participants exceeded this threshold, the corresponding frames were considered as moments of JVA.


Following the identification of JVA moments, we analyzed the dynamic attention characteristics of each participant during these periods using the ambient-focal attention $\mathcal{K}$. To compute coefficient $\mathcal{K}$ for each participant, we used the Real-Time Advanced Eye Movements Analysis Pipeline (RAEMAP) \cite{RAEMAP}, an eye movement processing library. The coefficient $\mathcal{K}$ is a dynamic indicator that captures the fluctuation between ambient and focal visual search behaviors \cite{krejtz2016discerning} (see Equation \ref{eq:coefficient_k}).


\begin{equation}
    \mathcal{K}_i = \dfrac{d_i - \mu_d}{\sigma_d} - \dfrac{a_{i+1} - \mu_a}{\sigma_a} 
    \label{eq:coefficient_k}
\end{equation}

Where $\mu_d$, $\mu_a$ are the mean fixation duration and saccade amplitude, respectively, and $\sigma_d$, $\sigma_a$ the standard deviation of the fixation duration and saccade amplitude respectively, which is then computed over all n fixations. A positive $\mathcal{K}$ value indicates a more focal scanning pattern, while a negative value reflects a more ambient scanning strategy during joint attention.

\section{Results}

\subsection{Percentages of JVA}

To quantify the prevalence of JVA throughout the interaction sessions, we calculated the percentage of JVA as a ratio between the number of frames filtered as JVA moments (those exceeding our established similarity threshold of 0.7) and the total number of frames in videos of each session. This approach provides a measure of joint attention frequency that allows meaningful comparisons across varying activities and contexts. Our method of quantifying JVA is nearly related to the technique introduced by Schneider et al. \cite{schneider-2021-shared-gaze-visualization}, which used the number of overlapping gaze positions during a time window over the total gaze points during the entire activity to quantify the JVA in a collaborative setting. By adapting this approach to egocentric data, our measure captures the proportion of shared attention during interactions.
As shown in Table \ref{tab:JVA_percentages}, our methods captured the highest level of joint attention in A4 and A5, where the participants interacted with a shared object, while the lowest was in A2 and A3, where the participants performed most tasks independently. While our observation proves the intuitive idea of the level of collaboration in each task, it also establishes the potential utility of our approach in joint attention detection.

\begin{table}[h!]
\centering
\caption{Percentage of joint visual attention in different everyday activities}
    \begin{tabular}{|l|l|c|}\hline
    
    \textbf{Activity}                   & \textbf{Percentage of JVA} \\\hline
    
    \textbf{A1:} Playing a game         & 23.74\% \\
    \textbf{A2:} Making coffee          & 3.97\% \\
    \textbf{A3:} Cooking and eating     & 5.12\% \\
    \textbf{A4:} Watching a video on mobile & 46.44\% \\
    \textbf{A5:} Watching Television    & 44.16\% \\\hline
    
    \end{tabular}
\label{tab:JVA_percentages}
\end{table}

\subsection{Dynamics of coefficient $\mathcal{K}$}

\begin{figure}
    \centering
    \includegraphics[width=\linewidth, height=20.5cm]{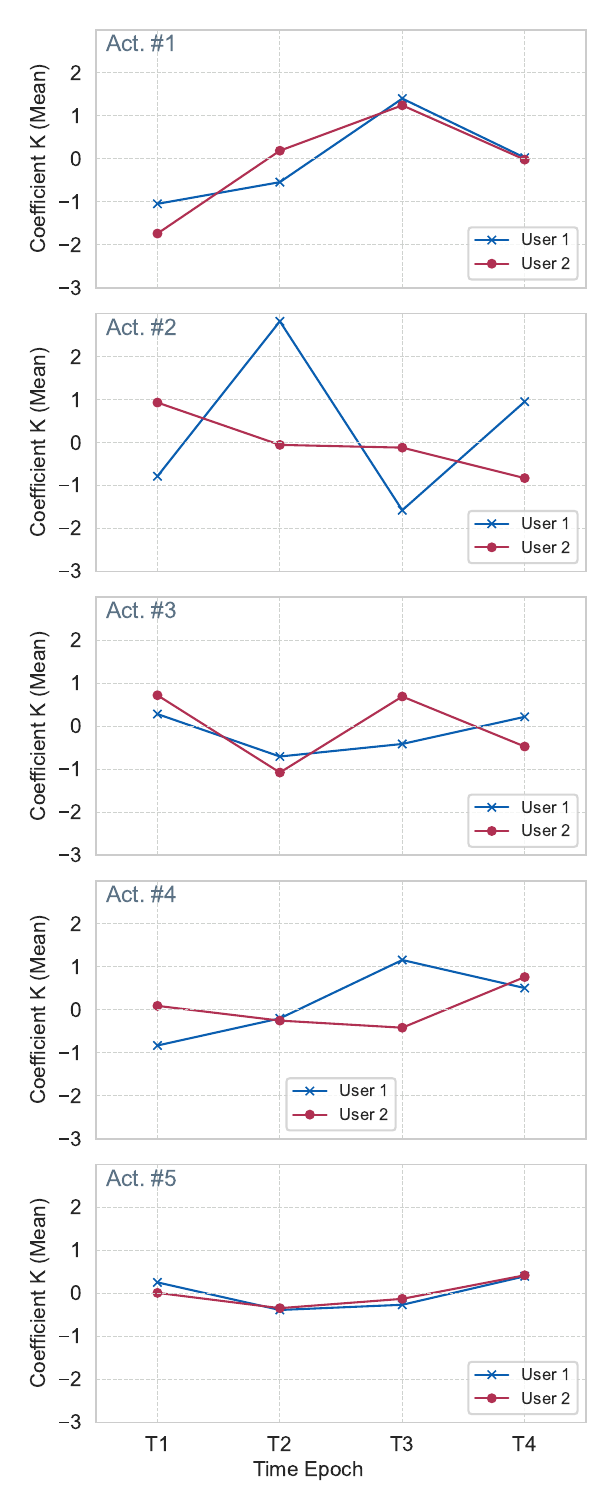} 
    \caption{Dynamics of ambient and focal attention using coefficient $\mathcal{K}$ between users in each session across four time-epochs.}
    \label{fig:dynamic-k}
\end{figure}

We analyzed the dynamics of the ambient-focal attention coefficient $\mathcal{K}$ to investigate changes in participants' attention patterns during periods of joint visual attention. Our temporal analysis divided each experiment into four epochs and computed the attentional characteristics. For this purpose, we employed the coefficient $\mathcal{K}$ as a quantitative measure of visual attention, where negative readings indicate ambient attention, while positive values indicate focal attention. Our examination of $\mathcal{K}$ values revealed changes in individual attention patterns across four temporal segments.

For a macro-level analysis of the behaviors, we manually annotated the events occurring each time by a panel of three experts, assigning a high-level annotation to the overall event during each temporal window. Then, we compared the behavior of $\mathcal{K}$ with the events in each temporal window for a qualitative study (see Table \ref{tab:k_dynamics}). Among instances where participants interact with the same shared object, we observed the $\mathcal{K}$ to converge between participants (A1:T3-T4, A4:T4, A5:T1-T4). Moreover, we observed dissimilar $\mathcal{K}$ readings where interactions were independent (A2:T1-T4) or when attention spread on different objects (A1:T1-T2). Our results suggest the potential of $\mathcal{K}$ for measuring joint attention dynamics when combined with environmental events.

\begin{table*}[ht]
\centering
\caption{High-level event annotations for time epochs in joint attention study.}
\begin{tabular}{|l|c|c|c|c|}
\hline
\textbf{Activity} & \multicolumn{4}{c|}{\textbf{Epoch Annotation}} \\
\cline{2-5}
                  & \textbf{Epoch 1} & \textbf{Epoch 2} & \textbf{Epoch 3} & \textbf{Epoch 4} \\
\hline
\textbf{A1:} Playing a game             & Preparing      & Preparing      & Playing       & Playing \\
\textbf{A2:} Making coffee              & Grabbing items & Grabbing items & Pouring       & Drinking \\
\textbf{A3:} Cooking and eating         & Preparing      & Preparing      & Eating        & Eating \\
\textbf{A4:} Watching a video on mobile & Walking        & Watching       & Watching      & Conversation, Watching \\
\textbf{A5:} Watching Television        & Watching       & Watching       & Watching      & Watching \\
\hline
\end{tabular}
\label{tab:k_dynamics}
\end{table*}


\section{Discussion}

The popularity of smart wearable devices, including AR/VR headsets and head-mounted cameras, has significantly advanced the collection and analysis of egocentric data for social attention research \cite{abeysinghe2025framework}. These technologies enable researchers to capture first-person perspectives complete with eye tracking, head movement, and environmental context during natural interactions. This methodological approach offers advantages for studying JVA

Our methodology extends previous work by adding another step to the identification of joint attention moments by incorporating analysis of attention patterns exhibited during these socially significant interactions. We use the ambient-focal attention with coefficient $\mathcal{K}$ as an analytical measure to explore the potential of advanced gaze metrics to study JVA which will be a novel contribution to the field. 


 Our study encountered several technical challenges inherent to egocentric data analysis. The dynamic nature of the visual field in egocentric recordings presents substantial processing difficulties, as participant movements constantly alter the frame of reference. This is particularly evident in the Table \ref{tab:JVA_percentages}, in Activities 2 and 3. Additionally, the relatively low resolution of Project Aria glasses decreases the critical details, and distinguishing features become obscured, reducing the discriminative power of similarity metrics. Varying lighting conditions across different recording environments also affected our similarity calculations and ultimately affected in detecting frames with JVA. 

Looking ahead, we plan to enhance the JVA identification component of our methodology by incorporating emerging techniques like Segment Anything \cite{metaIntroducingSegment} model, an advanced image segmentation algorithm by Meta. By detecting specific objects rather than relying solely on visual similarity, we may overcome some of the limitations imposed by lighting variations and dynamic viewpoints. 

The introduction of the ambient-focal analysis into JVA research opens new research questions about the relationship between attention patterns and effective collaboration. Future studies could investigate whether certain patterns of ambient-focal coefficient  $\mathcal{K}$ dynamics correlate with more successful collaborative outcomes or more efficient task completion. We plan to expand this utility study on advanced eye tracking metrics to incorporate other advanced gaze measures like gaze transition entropy \cite{cui2024gaze}. Ultimately, our study lays the groundwork for future research on JVA with detailed interpretations. 

\section{Conclusion}

This study highlights the value of analyzing joint visual attention (JVA) through egocentric data in natural settings. Our methodology combines spatiotemporal tube analysis with advanced gaze metrics to identify and characterize shared attention moments. Results show JVA patterns vary by task type, with higher rates during collaborative object-focused activities compared to independent tasks. The convergence of ambient-focal attention coefficients during shared object interaction suggests attentional synchronization that may enhance collaboration. These insights provide a detailed understanding of visual attention coordination in everyday activities. Our approach offers researchers a tool for investigating social attention in ecologically valid contexts, with applications in developmental psychology, human-computer interaction, and social robotics.

\bibliographystyle{IEEEtran}
\balance
\bibliography{ref}

\end{document}